\documentclass[twocolumn,showpacs,preprintnumbers,amsmath,amssymb]{revtex4-1}
\usepackage{comment} 
\usepackage{graphicx}
\usepackage{dcolumn}
\usepackage{bm}
\usepackage[compact]{titlesec}
\usepackage{soul}
\usepackage{ulem}
\usepackage{tensor}

\usepackage{xcolor}

\usepackage{hyperref}
\newcommand{\rmd}{{\rm d}}
\newcommand{\rh}{r_{h}}

\begin{document}

\preprint{CTPU-PTC-22-09}
\title{Rectifying No-Hair Theorems in Gauss-Bonnet theory}

\author{Alexandros Papageorgiou}
\email{papageo@ibs.re.kr}

\author{Chan Park}
\email{iamparkchan@gmail.com}

\author{Miok Park}
\email{miokpark76@ibs.re.kr}

\affiliation{Center for Theoretical Physics of the Universe, IBS, 34126 Daejeon, South Korea}

\begin{abstract}
\noindent  
We revisit the no-hair theorems in Einstein-Scalar-Gauss-Bonnet theory with a general coupling function between the scalar and the Gauss-Bonnet term in four dimensional spacetime. In the case of the old no-hair theorem the surface term has so far been ignored, but this plays a crucial role when the coupling function does not vanish at infinity and the scalar field admits a power expansion with respect to the inverse  of the radial coordinate in that regime. We also clarify that the novel no-hair theorem is always evaded for regular black hole solutions without any restrictions as long as the regularity conditions are satisfied. 

\end{abstract}

                             
\maketitle

\section{Introduction}

The uniqueness theorems {\cite{Israel:1967wq,Israel:1967za}} made us believe that black holes might have no hair except for the mass, electromagnetic charge or angular momentum. This motivated the assertion of the no-hair theorem \cite{Chase1970,Bekenstein:1972ny,Teitelboim:1972pk} which proved the non-existence of black hole solutions with non-trivial scalar field for asymptotically flat spacetime. The cases for the massive vector or spinor field were also discussed respectively in \cite{Bekenstein1972,Hartle1971}. However the desire to find new black hole solutions led to the discovery of several kinds of non-trivial field such as colored black holes \cite{Bizon1990} or Skyrmion hair black holes \cite{Luckock1986}. More recently, the evasion of the no-hair theorem has been shown for Einstein theory with a Gauss-Bonnet term which couples to massless scalar fields \cite{Kanti:1995vq,Sotiriou:2013qea,Sotiriou:2014pfa,Antoniou:2017acq,Antoniou:2017hxj,Silva:2017uqg,Doneva:2017bvd,Doneva:2021tvn}. These results were subsequently extended to consider self-interactions \cite{Bakopoulos:2020dfg,Macedo:2019sem} or a cosmological constant \cite{Bakopoulos:2018nui,Lee:2021uis}. 

The theory of general relativity with higher derivatives is well motivated. The early study arose in quantum field theory by finding that higher derivative terms stabilize the divergent structure of gravity and helps to establish a renormalizable theory of gravity in the absence of matter fields \cite{tHooft:1974toh, Stelle:1978}. Inspired by this result, application to cosmology was initially investigated in \cite{Starobinsky:1980} and later more broad construction of modifying Einstein gravity has been widely studied in many works (see \cite{Clifton:2011jh} and references therein). Furthermore, from the perspective of string theory, taking the low energy limit, gravity theory is reduced to Einstein theory with higher derivative terms whose coefficient is $\alpha'$, the inverse string tension, and associated with the dilaton coupling. Therefore the $\alpha'$ correction is considered as the stringy effect beyond Einstein gravity. These circumstances draw attention to the existence of black hole solutions in higher derivative theories. 

In particular, the Gauss-Bonnet theory has a special interest since it is topological in four dimensions. With this virtue of the theory, the no-hair theorems for Einstein-Scalar-Gauss-Bonnet theory (ESGB) are argued in \cite{Antoniou:2017acq} but their analysis was not complete both for the old no-hair theorem as well as the novel version of the theorem as follows. 

Firstly, the old no-hair theorem for ESGB theory in \cite{Antoniou:2017acq} follows the same steps as in the original argument by Bekenstein \cite{Bekenstein:1972ny}. They start with the equation of motion for a scalar field $\varphi(r)$ multiplied by the coupling function $f(\varphi)$ and take an integration over spacetime. Then performing an integration by parts the surface term is dropped and only the bulk term remains. This leads to the condition $f(\varphi) > 0$ as a necessary requirement for the evasion of the old no-hair theorem. In the original work of Bekenstein \cite{Bekenstein:1972ny}, the surface term vanishes because either the field is massive and therefore enjoys a Yukawa-like, exponential decay, or the field is massless and the theory is shift-symmetric. In the latter case, the surface term is unphysical since it can always be cancelled by a field shift which leaves the Lagrangian invariant. On the other hand, in ESGB theory with a massless scalar field, the fall off behaviour of the scalar field at infinity becomes slower such that the surface term survives in general and it remains physical when the theory does not enjoy a shift symmetry. Thus the no-hair theorem should be understood on this basis. We rectify the assumptions for the no-hair theorem in this paper and justify the existence of black hole solutions for $f(\varphi)<0$.  

Secondly, the novel no-hair theorem involves analyzing the asymptotic behaviour of the energy-momentum tensor near the horizon and at infinity under the assumption of a regular black hole solution and determines the possibility that the energy-momentum tensor smoothly matches both asymptotic limits. The application to the ESGB theory in \cite{Antoniou:2017acq} shows that imposing the regularity condition on the horizon is necessary to guarantee the evasion of the novel no-hair theorem and to further ensure the evasion the authors require the derivative of the energy-momentum tensor $(T_{r}{}^{r})'$ to be negative near the horizon. We revisit this and find that the later condition plays no role in determining whether physically acceptable solutions exist.

In this paper, we revisit the no-hair theorem in ESGB theory studied in \cite{Antoniou:2017acq} and revise their argument with respect to the old no-hair theorem as well as extend the analysis of the novel no-hair theorem. In section 3 we point out the omission of the surface term for the old no-hair theorem in \cite{Antoniou:2017acq}. In section 4 we show that in the case of the novel theorem the energy-momentum tensor can be regular without any further constraints as long as the condition for regularity of the scalar field is satisfied. Then we demonstrate our argument with numerical solutions in section 5. Thus we clarify the conditions for the no-hair theorem to hold and conclude that the no-hair theorem is more easily evaded than previously studied.  
 
\section{Einstein-Scalar-Gauss-Bonnet Theory}

We start with the gravity action as follows
\begin{align}
S &=  \int \rmd^4 x \sqrt{-g} \bigg[\frac{R}{2\kappa^2} -\frac{1}{2} \nabla_{\alpha} \varphi \nabla^{\alpha} \varphi + f(\varphi) \mathcal{G}\bigg] 
\end{align}
where we set $2 \kappa^2 = 16 \pi G =1$ and $\mathcal{G}$ denotes the Gauss-Bonnet term that is written as
\begin{align}
\mathcal{G} = R_{\mu \nu \rho \sigma}R^{\mu \nu \rho \sigma} - 4 R_{\mu \nu} R^{\mu \nu} +R^2
\end{align}
and which leads to the Einstein equations as follows
\begin{align}
R_{\mu \nu} &- \frac{1}{2}R g_{\mu \nu} = T_{\mu \nu} = \kappa^2 \bigg[ \partial_{\mu} \varphi \partial_{\nu} \varphi - \frac{1}{2} g_{\mu \nu} \partial_{\rho} \varphi \partial^{\rho} \varphi  \nonumber\\ 
&-  (g_{\rho \mu} g_{\lambda \nu} + g_{\lambda \mu} g_{\rho \nu}) \eta^{\kappa \lambda \alpha \beta} \tilde{R}^{\rho \gamma}{}_{
\alpha \beta} \nabla_{\gamma} \nabla_{\kappa} f \bigg]
\end{align}
where $\tilde{R}^{\rho \gamma}{}_{\alpha \beta} = \eta^{\rho \gamma \sigma \tau} R_{\sigma \tau \alpha \beta} = \frac{\epsilon^{\rho \gamma \sigma \tau}}{\sqrt{-g}} R_{\sigma \tau \alpha \beta}$
and the scalar field equation is
\begin{align}
\nabla^{2} \varphi + \dot{f} \mathcal{G} = 0
\end{align}
where `` $\dot{}$ " indicates the variation with respect to the scalar field $\varphi$. Employing the following metric ansatz
\begin{align}
\textrm{d}s^2 =-A(r) \textrm{d}t^2 + \frac{1}{B(r)}\textrm{d}r^2 + r^2 \textrm{d} \Omega_2,
\end{align}
the equations of motions are written as
\begin{align}
&\frac{r B'+B-1}{r^2 B}+\frac{\kappa ^2}{2}  \varphi '^2 + \frac{4 \kappa ^2}{r^2 B} \bigg[ \nonumber\\
&(1-3 B) B' \dot{f} \varphi '-2 (B-1) B \left(\ddot{f} \varphi '^2+\dot{f} \varphi ''\right)\bigg] = 0,\label{eq:tt} \\
&\frac{A'}{A r}+\frac{B-1}{B r^2}-\frac{\kappa ^2}{2}  \varphi '^2 + \frac{4 \kappa ^2 (1-3 B)  A' \varphi ' \dot{f}}{A r^2}= 0,\label{eq:rr} \\
&\frac{A' \left(2 A-r A'\right)}{4 A^2 r} +\frac{B' \left(r A'+2 A\right)}{4 A B r}+\frac{A''}{2 A} +\frac{\kappa ^2}{2} \varphi '^2  \nonumber\\
& - \frac{2 \kappa^2}{Ar} \bigg[ \dot{f} \varphi' \bigg(2 B A''+3 A' B'-\frac{B A'^2}{A} \bigg) \nonumber\\
&+ 2 B A' \left(\ddot{f}\varphi '^2+ \dot{f} \varphi ''\right) \bigg] = 0,\label{eq:phiphi}\\
&\varphi'' + \frac{1}{2} \varphi ' \left(\frac{A'}{A}+\frac{B'}{B}+\frac{4}{r}\right) + \frac{2 \dot{f}}{A r^2} \bigg[\frac{(3 B-1) A' B'}{B} \nonumber\\
&-\frac{(B-1)}{A} \left(A'^2-2 A A''\right)\bigg] = 0\label{eq:scalar}
\end{align}
where `` $'$ " indicates the variation with respect to the radial coordinate $r$. 

If we assume the existence of a regular black hole, we require the following boundary conditions near the horizon
\begin{align}
&A(r) \sim A_h \epsilon , \; \; \; B(r) \sim B_h \epsilon, \; \; \; \varphi(r) \sim \varphi_h + {\varphi_{h,1}} \epsilon  \label{eq:NHexp}  
\end{align}
where $\epsilon = r - r_h$ is the expansion parameter and $\varphi_h$ is a finite value near the black hole horizon. Firstly, as was pointed out in several works before \cite{Kanti:1995vq,Antoniou:2017acq}, in order to ensure that the scalar field and its derivatives are finite one requires the following constraint, valid on the horizon
\begin{align}
    {\varphi_{h,1}} &= - \frac{\rh}{4\dot{f}_h}\left(1\mp \sqrt{1 - \frac{96}{\rh^4} \dot{f}_h^2}\right), \label{eq:dvarphi}\\
    B_h&= \frac{2}{\rh}\left(1 \pm \sqrt{1 -\frac{96}{\rh^{4}} \dot{f}_h^2}\right)^{-1}\label{eq:Bh}
\end{align}
where $\dot{f}_h = \dot{f}(\varphi_h)$. We found that the numerical solutions are generated only for the minus sign in front of root in (\ref{eq:dvarphi}) with the plus sign in (\ref{eq:Bh}) and will just consider this case hereafter. To avoid $\varphi''(r_h)$ being divergent the inside of the root should not be zero, namely 
\begin{align}
\dot{f}_{h}^{2}< \frac{r_h^4}{96}. \label{eq:constraint2}
\end{align}
These regularity conditions (\ref{eq:dvarphi})-(\ref{eq:constraint2}) ensure that the solutions correspond to a regular black hole spacetime. We also write the near horizon expansion of the Riemann scalar invariant
\begin{align}
    R_{\alpha \beta \mu \nu} R^{\alpha \beta \mu \nu}&\sim \frac{64{\left(2\pm\sqrt{1-\frac{96}{\rh^{4}}\dot{f}_{h}^{2}}-\frac{24}{\rh^{4}}\dot{f}_{h}^{2}\right)} {\left(1-\frac{24}{\rh^{4}}\dot{f}_{h}^{2}\right)}}{\rh^{4}\left(1\pm\sqrt{1-\frac{96}{\rh^{4}}\dot{f}_{h}^{2}}\right)^{4} }
\end{align}
which is finite unless the size of the horizon becomes zero $r_h \to 0$. This ensures that the spacetime is not a naked singularity but a regular black hole. 

At infinity, the asymptotic flatness requires that the metric and scalar field are found to be
	\begin{align}
	&A(r) \sim 1+\frac{A_1}{r}, \; B(r) \sim 1+\frac{A_1}{r}, \; \varphi(r) \sim \varphi_{\infty } +\frac{\varphi_1}{r} \label{eq:vpinf}
	\end{align}
where $\varphi_{\infty}$ takes a finite value that is physical if the theory does not enjoy a shift symmetry on $\varphi$. Here $\varphi_1$  is deeply related to scalar charge \cite{Gibbons:1987ps}, which is defined by
\begin{align}
Q = -\frac{1}{4 \pi} \int_{S^2} \textrm{d}^2 \Sigma^{\mu} \nabla_{\mu} \varphi
\end{align}
at infinity. Plugging the expansion of the metric into the Gauss-Bonnet term yields
\begin{align}
    \mathcal{G}&\sim \frac{48}{\rh^{4}}\left(1\pm\sqrt{1-\frac{96}{\rh^{4}} \dot{f}_{h}^2}\right)^{-2} + \mathcal{O}(\epsilon), \; (r \to r_h) \\
    \mathcal{G}&\sim \frac{12 A_1^2}{r^6} + \mathcal{O}(r^{-7}),  \; \; (r \to \infty)
\end{align}
which are positive at both asymptotics. Using these expansions, we examine the existence of black hole solutions.

\section{Old no-hair theorem}

As studied in \cite{Kanti:1995vq,Antoniou:2017acq}, let us start with the scalar field equation multiplied by the coupling function $f(\varphi)$ and then take an integration over four dimensional spacetime 
\begin{align}
&\int_{\mathcal{V}} \textrm{d}^4 x \sqrt{-g}f[\nabla^{2}  \varphi + \dot{f} \mathcal{G}]  =0 \\
&=  - \int_{\mathcal{V}} \textrm{d}^4 x \sqrt{-g} \dot{f} \bigg( \partial^{\mu} \varphi\partial_{\mu} \varphi - f \mathcal{G} \bigg) \nonumber\\
& + \int_{\partial \mathcal{V}} \textrm{d}^3 x \sqrt{-h} f n^{\mu}\partial_{\mu} \varphi
\end{align}
where $h$ is the induced metric on a hypersurface defined by a normal vector $n^{\mu}$. Since we assume that the static scalar field depends on the radial coordinate, only the $\mu=r$ component makes a non-trivial contribution. Factoring out the time and angular integrations, it reduces to
	\begin{multline}
	\int^{\infty}_{r_h} \textrm{d} r \sqrt{\frac{A}{B}} r^2 \dot{f}  \bigg(\partial^{\mu} \varphi \partial_{\mu} \varphi - f \mathcal{G} \bigg)  \\
	- \bigg(\sqrt{\frac{A}{B}} r^2 g^{rr} f \partial_{r} \varphi  \bigg) \bigg|_{r \to \infty} = 0 \label{eq:oldNoH}
	\end{multline}
where the surface term, that is the second line above, vanishes at the horizon since $g^{rr} \to 0$ as $r \to r_h$. Substituting the asymptotic expansion (\ref{eq:vpinf}) to the surface term, we see that it remains finite and approaches $f(\varphi_{\infty}) \varphi_1$. The presence of a surface term makes it non-trivial to prove the non-existence of black hole solutions for a general coupling function $f(\varphi)$ and it is rather suggestive that the no-hair theorem is evaded. But for the special cases such as in the absence of the surface term by $f(\varphi_{\infty}) =0$ or $\varphi_1=0$, the no-hair theorem still holds when $f(\varphi)<0$, but is evaded again if $f(\varphi)>0$ as previously studied in \cite{Antoniou:2017acq}. Nevertheless, the case of $f(\varphi_{\infty}) =0$ cannot be achieved for a general coupling. For the cases of $f(\varphi) = \alpha e^{\gamma \varphi(r)}$ or $f(\varphi) = \alpha \varphi(r)^{-n}$ with positive $n$, the regular scalar field solution cannot make $f(\varphi_{\infty})$ to vanish and hence the surface term is always present unless $\varphi_1=0$. On the other hand, for the cases of $f(\varphi) = \alpha \varphi(r)^n$ or $f(\varphi) = \alpha(1-e^{\gamma \varphi^2})$, the surface term will disappear if $\varphi_{\infty} = 0$ and so the no-hair theorem still holds for $f(\varphi)<0$, while if $\varphi_{\infty} \neq 0$ the surface term survives. For the later case, black hole solutions would exist for either positive or negative sign of a coupling function. Thus, black hole solutions exist regardless of the sign of $f(\varphi)$ in these cases. This fact will be numerically demonstrated with the explicit coupling functions in section V.

\section{Novel no-hair theorem}

The novel no-hair theorem was formulated for scalar fields that are minimally coupled to gravity in \cite{Bekenstein:1995un}. They assume the positivity of the energy density $\mathcal{E}=-\tensor{T}{_{t}^{t}}>0$ and illustrate the asymptotic behaviours of $T_r{}^{r}$ and $(T_r{}^{r})'$ at horizon and infinity. They then show the impossibility of smoothly matching these asymptotic conditions by making explicit use of the Einstein equations. The same methodology was applied to the non-minimally coupled ESGB theory, in \cite{Kanti:1995vq,Antoniou:2017acq}, which showed that the smooth matching of the energy-momentum tensor at both asymptotics may be achieved if $\tensor{T}{_{r}^{r}}>0$ and $(T_{r}{}^{r})'<0$ close to the horizon. We find that the second assumption is not required to smoothly connect the energy-momentum tensor at both asymptotics. Our argument takes the following form. 

\begin{figure}
    \centering
    
    \includegraphics[scale=0.35]{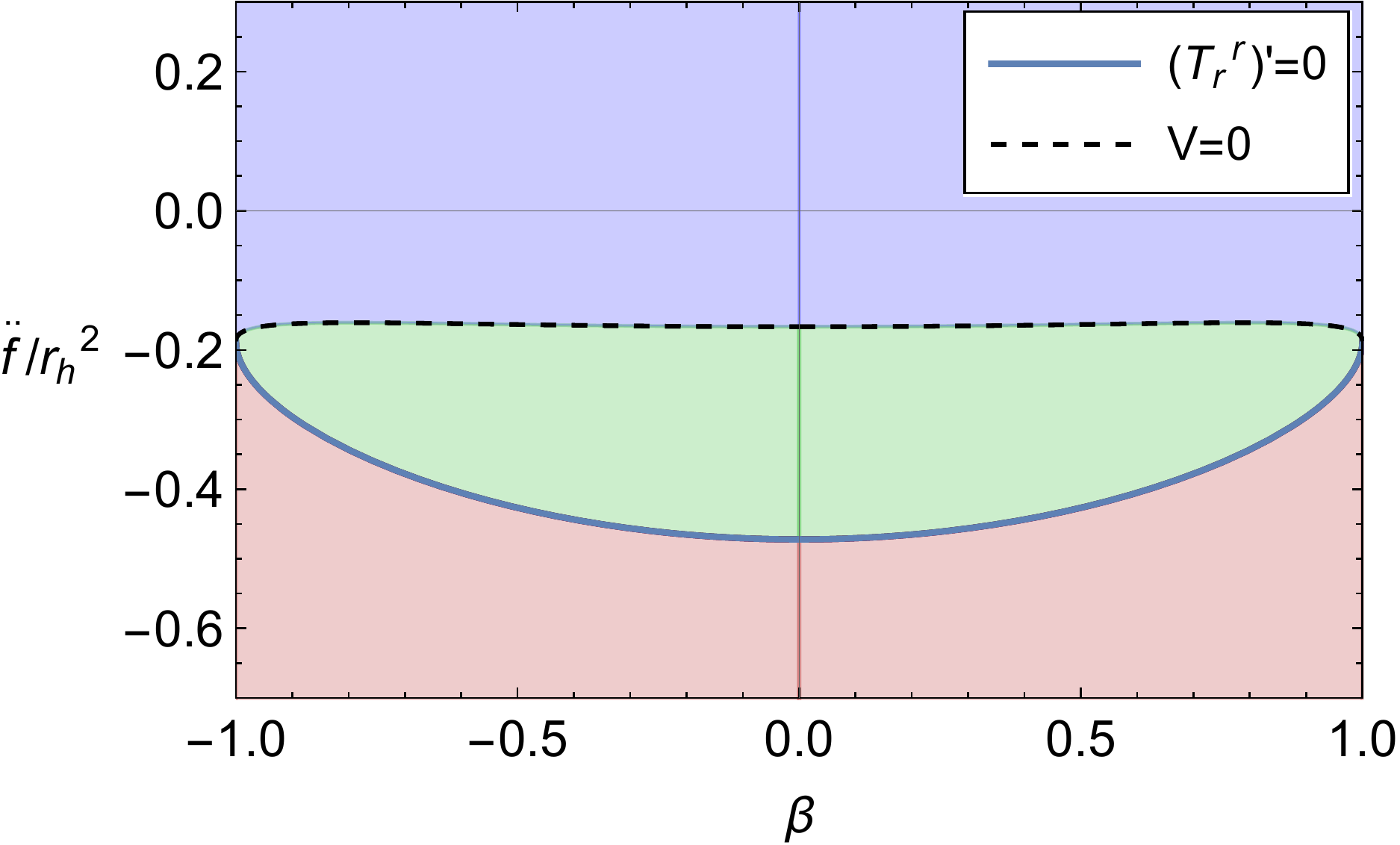}
    
    \caption{Parameter space for $(T_{r}{}^r)'|_{r \to r_h}$.  The blue $(V>0)$ and green $(V<0)$ regions satisfy $(T_r{}^r)'|_{r \to {r_h}}<0$ while the red region $(T_r{}^r)'|_{r \to r_h}>0$, and $\beta = \pm 1$ is excluded.}
    \label{fig:novel}
\end{figure}

We first expand the energy momentum tensor and its derivative around the horizon as well as at infinity
\begin{align}
T_{r}{}^{r}&= T_{t}{}^{t}= -\frac{1}{r B'} T_{\theta}{}^{\theta} = - \frac{2B'\dot{f} \varphi'}{r^2} +{\cal O}(\epsilon), \\
(T_r{}^{r})'&= \frac{BA'}{A} \bigg[\frac{4 \left(r B'+1\right) \dot{f} \varphi '}{r^3}-\frac{r \varphi '^2}{4 (r + 2 \dot{f} \varphi')} \nonumber\\
&\; \; \; \; \; -\frac{2(\ddot{f} \varphi '^2+\dot{f} \varphi '')}{r (r + 2 \dot{f} \varphi')}\bigg]+{\cal O}(\epsilon)
\label{eq:Trrp}
\end{align}
and
\begin{align}
&T_{r}{}^{r}= - T_{t}{}^{t} =- T_{\theta}{}^{\theta}= \frac{1}{4} \varphi'^2 + \mathcal{O}(r^{-5}) \label{eq:TrrNH}\\
&(T_{r}{}^{r})'= -\frac{1}{r} \varphi'^2 + \mathcal{O}(r^{-6}) \label{eq:TrrInf}
\end{align}
where the Gauss-Bonnet term is subdominant at large $r$. The product $\dot{f}\varphi'$ is always negative definite for regular black hole (\ref{eq:dvarphi}) and results in
\begin{align}
&T_r{}^{r}|_{r \to r_h} > 0, \qquad (T_r{}^{r})'|_{r \to r_h} : \textrm{undetermined}  \label{eq:SignTrrNH}\\
&T_r{}^{r}|_{r \to \infty} > 0, \qquad (T_r{}^{r})'|_{r \to \infty} <0  \label{eq:SignTrrINF}
\end{align}
where $(T_r{}^{r})'|_{r \to \infty}$ decays asymptotically as $r^{-5}$. As pointed out in \cite{Kanti:1995vq,Antoniou:2017acq}, this indicates that the energy density is negative near the horizon, $\mathcal{E} = - T_{t}{}^{t} < 0$, which is opposed to the minimally coupled case in \cite{Bekenstein:1995un}. This is an effect crucially driven by the Gauss-Bonnet term. The authors considered the possible smooth matching of the energy-momentum tensor at both asymptotics by requiring the condition
\begin{align}
V = \ddot{f} \varphi'^2 + \dot{f} \varphi'' = \partial_r \left(\dot{f}\varphi'\right)|_{r_h}>0
  \label{eq:V}
\end{align}
which guarantees $(T_r{}^{r})'|_{r \to r_h}$ to be negative in smoothly connecting the two asymptotic limits. However this is not the only way of matching. We firstly argue that the condition (\ref{eq:V}) is overly  restrictive. To explicitly show this, we explore the parameter space of $(T_{r}{}^{r})'|_{r \to r_h}$ by using the near horizon expansion which takes the form
\begin{align}
(T_{r}{}^{r})' =& -\frac{\beta^2\left[68-41\beta^2+\sqrt{1-\beta^2}\left(68-9\beta^2\right)\right]}{4 r_h^3\sqrt{1-\beta^2}\left(1+\sqrt{1-\beta^2}\right)^4}\nonumber\\
&-\frac{36 \ddot{f_h}\beta^2}{r_h^5\sqrt{1-\beta^2}\left(1+\sqrt{1-\beta^2}\right)^3}+{\cal O}(\epsilon)
\end{align}
where $\ddot{f}_h$ denotes $\ddot{f}(\varphi_h)$ and we used a new variable
\begin{align}
\beta = \pm \frac{\sqrt{96} |\dot{f_h}|}{r_h^2}
\end{align}
which ranges $-1< \beta < 1$ and this will carry the same sign as $\alpha$ in our examples. We plot the boundary in which $(T_{r}{}^{r})'|_{r \to r_h}$ changes sign in Fig.\ref{fig:novel} as a solid line. Inspection of the figure allows one to visualize the various possibilities regarding the sign of $(T_r{}^{r})'$ near the horizon but this does not mean that all parameter space yields black hole solutions. As seen in Fig.\ref{fig:novel} the green region also makes $(T_r{}^{r})'$ to be negative even with a negative value of $V$. We secondly point out that in spite of the undetermination of $(T_r{}^{r})'$ the rest of the conditions in (\ref{eq:SignTrrNH}) -- (\ref{eq:SignTrrINF}) do not prevent having $(T_r{}^{r})'|_{r \to r_h} > 0$ to smoothly join the energy-momentum tensor from the horizon to infinity. Namely, either sign $(T_r{}^{r})'$ is allowed to match $T_r{}^{r}$ at both asymptotic regions. In conclusion the whole region except the points with $\beta=\pm1$ in Fig.\ref{fig:novel} should be considered to generate black hole solutions.

In the next section we explore all three regions and numerically find solutions for regular black holes with scalar hair for all three cases. 

\section{Examples}
\label{sec:example}

Here we verify the evasion of no-hair theorems for two specific coupling functions. In order to generate the numerical solution the expansion coefficients $A_h$, $\varphi_h$  are used as the parameters to give the initial conditions for equations (\ref{eq:tt})-(\ref{eq:scalar}), but $A_h$ is determined by the boundary condition at infinity. Here we produce a family of solutions for varying $\varphi_h$ fixing $r_h = 1$. 

\subsection[]{$f(\varphi) = \alpha e^{\gamma \varphi(r)}$}

\begin{figure}[b!]
\includegraphics[scale=0.265]{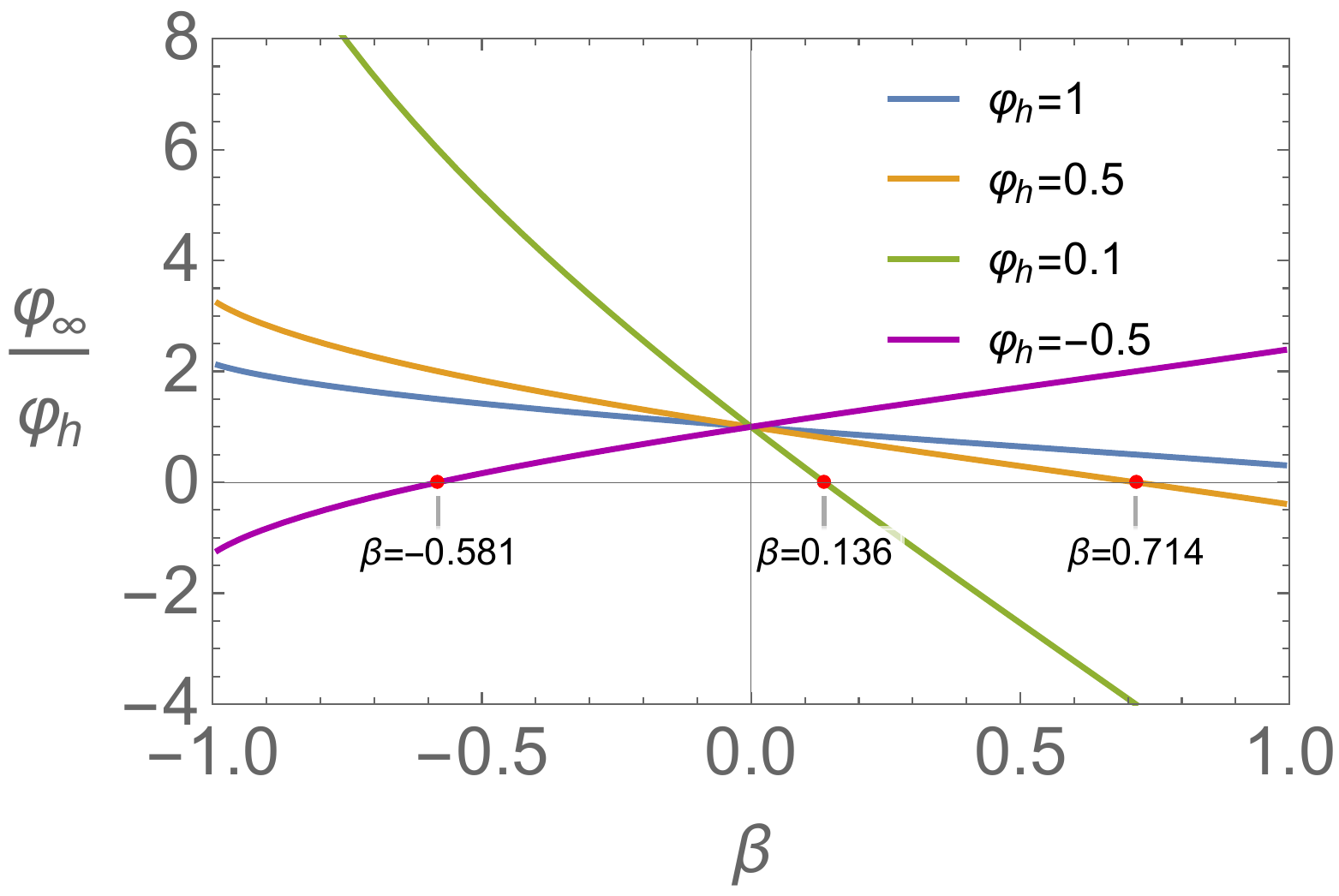} \includegraphics[scale=0.27]{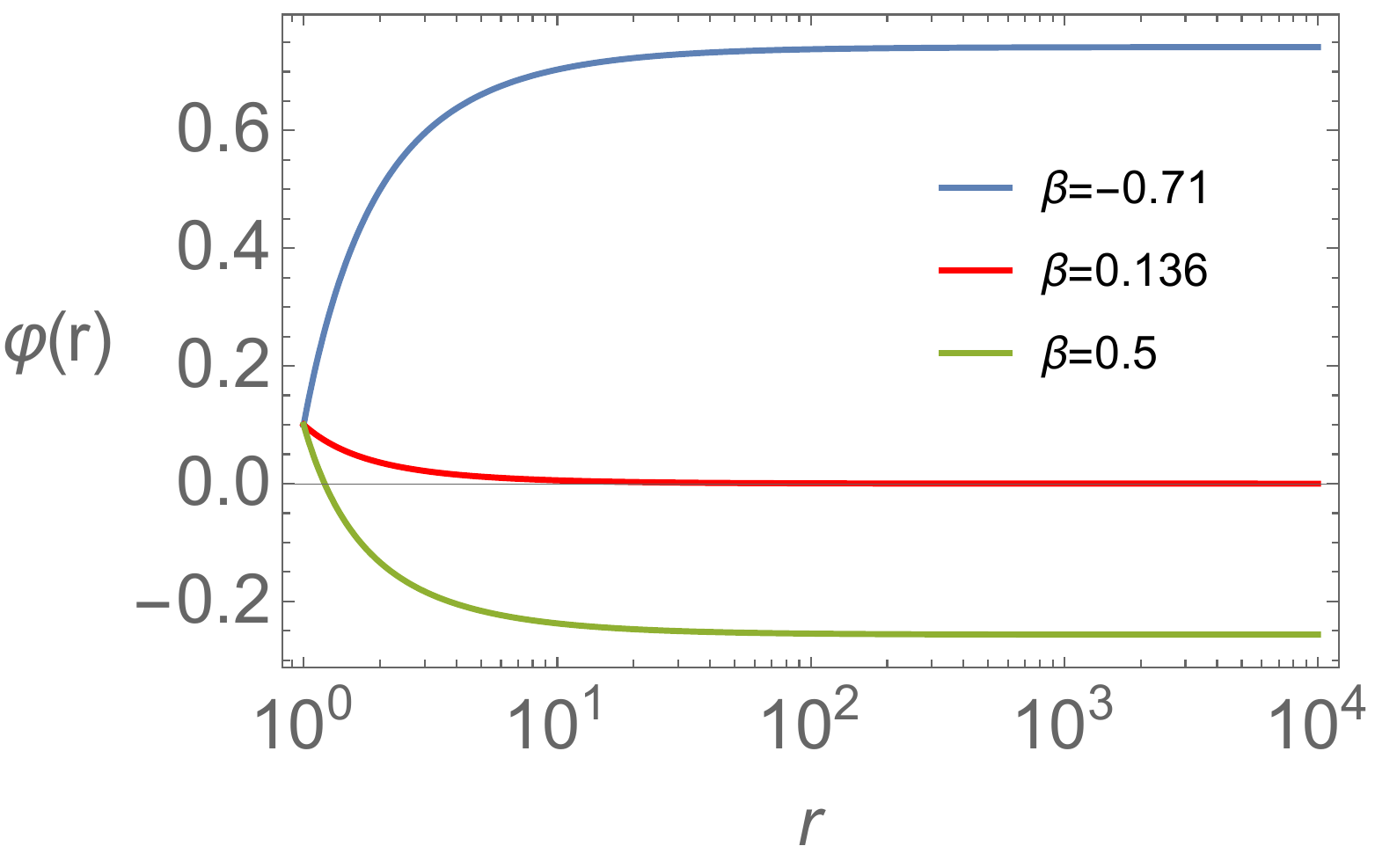} 
\caption{For $f=\alpha e^{\gamma \varphi}$  (left)  $\varphi_{\infty}/\varphi_h$ vs $\beta$ and (right) $\varphi(r)$ for different values of $\beta$ fixing $\varphi_h = 0.1$}\label{fig:ExAInf} 

\includegraphics[scale=0.252]{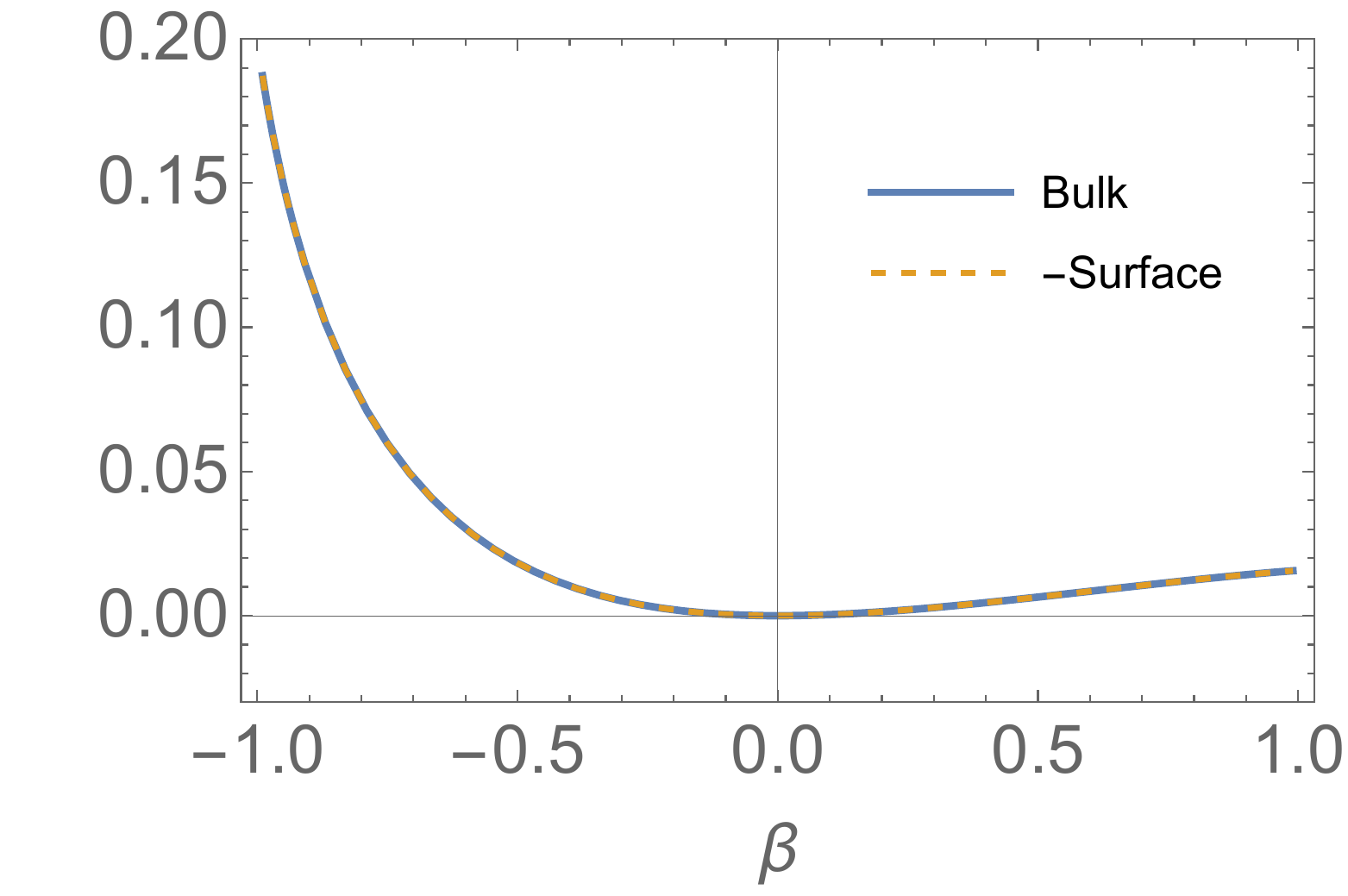} \includegraphics[scale=0.28]{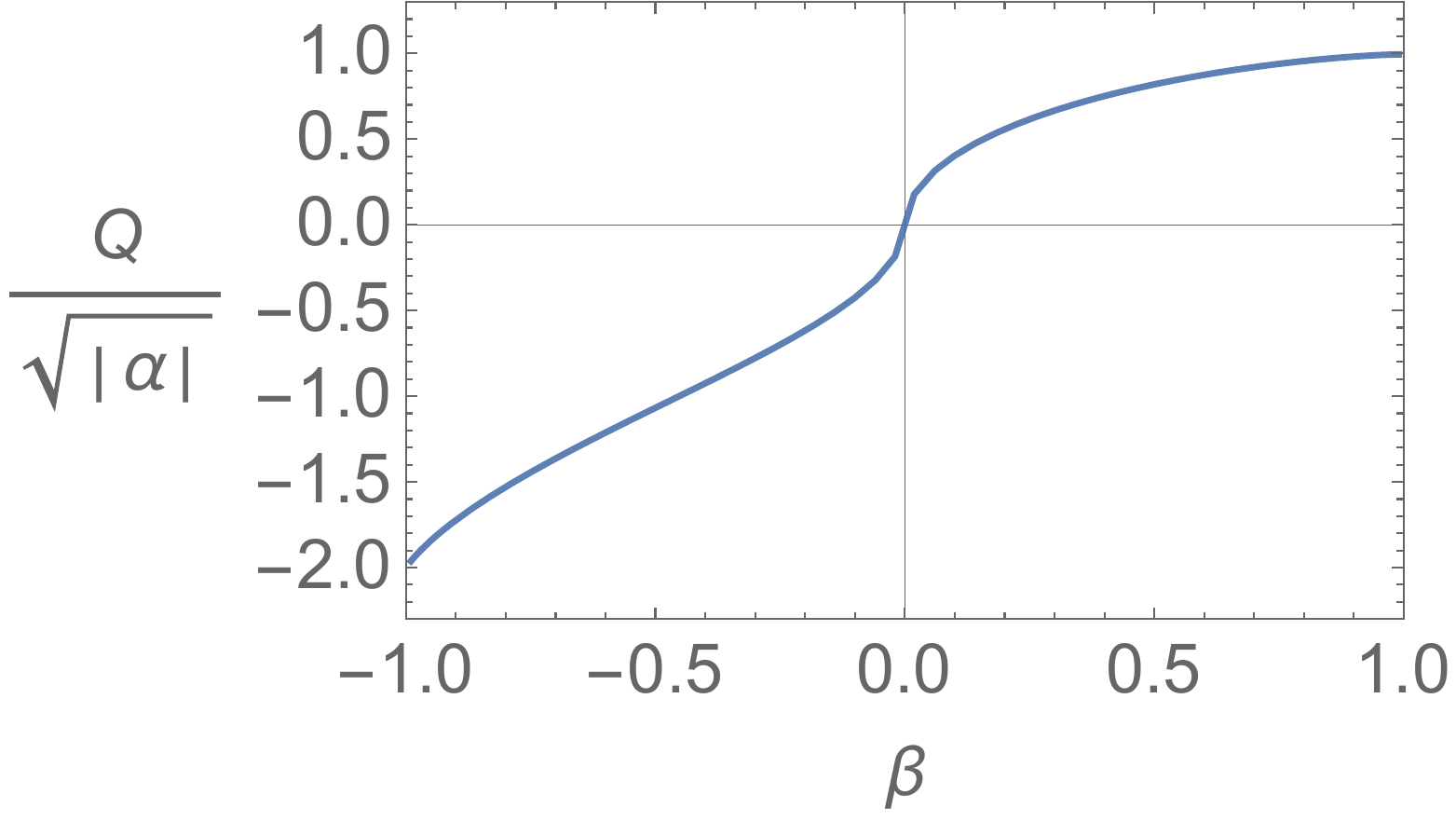} 
\caption{Old no-hair theorem: For $f=\alpha e^{\gamma \varphi}$ with $\varphi_h = 0.1$ (left) plot of bulk and surface term (right) the scalar charge $Q/{\sqrt{|\alpha|}}$ vs $\beta$ } \label{fig:ExAoldNH}
\end{figure}

As addressed before, the surface term does not disappear in this case unless $\varphi_1 =0$ and therefore the no-hair theorem is evaded for any values of $\alpha$ as long as the coupling function satisfies (\ref{eq:dvarphi})-(\ref{eq:Bh}). For a given $\varphi_h$, the value of $\varphi_{\infty}/\varphi_h$ is depicted as a function of $\beta$ and the scalar functions $\varphi(r)$ for the cases $\beta= -0.71$, $0.136$, and $0.5$ are displayed in Fig.\ref{fig:ExAInf}. We fix $\gamma=1$ and only focus on the positive case for $\gamma$ since the equations of motion are symmetric with  $-\gamma$ under the change of the sign of the scalar field $\varphi(r)$. Since there is a shift symmetry of the Dilaton field in this case, a non-zero value of $\varphi_{\infty}$ can be shifted to zero under a rescaling of the radial coordinate $r$ \cite{Kanti:1995vq}. To examine the evasion of the old no-hair theorem (\ref{eq:oldNoH}), we plot the bulk term and surface term, which are the first and second line in (\ref{eq:oldNoH}) respectively, as a function of $\beta$ in Fig.\ref{fig:ExAoldNH}. This demonstrates that the value of the surface term is not zero but takes the opposite sign to the bulk term and therefore plays a crucial role for the evasion of the old no-hair theorem. We also investigated the energy-momentum tensor, but since our parameters choices  ($\gamma=1$) are firmly in the blue region of Fig.\ref{fig:novel}, $T_{r}{}^{r}$ is positive and monotonically deceasing and $(T_{r}{}^{r})'$ is negative and monotonically increasing having positive $V$ in (\ref{eq:V}). We have found solutions for both the green as well as the red region of Fig.\ref{fig:novel} by changing the value of $\gamma$, but observed that the behavior of the metric functions differs considerably from the Schwarzschild-like behavior \cite{Magalhaes:2020sea}. We here exclude such exotic solutions. We instead will explore the green and red regions of Fig.\ref{fig:novel} for the novel no-hair theorem in the next example.

\subsection[]{$f(\varphi) = \alpha \varphi(r)^2$}

\begin{figure}[t!]
\includegraphics[scale=0.265]{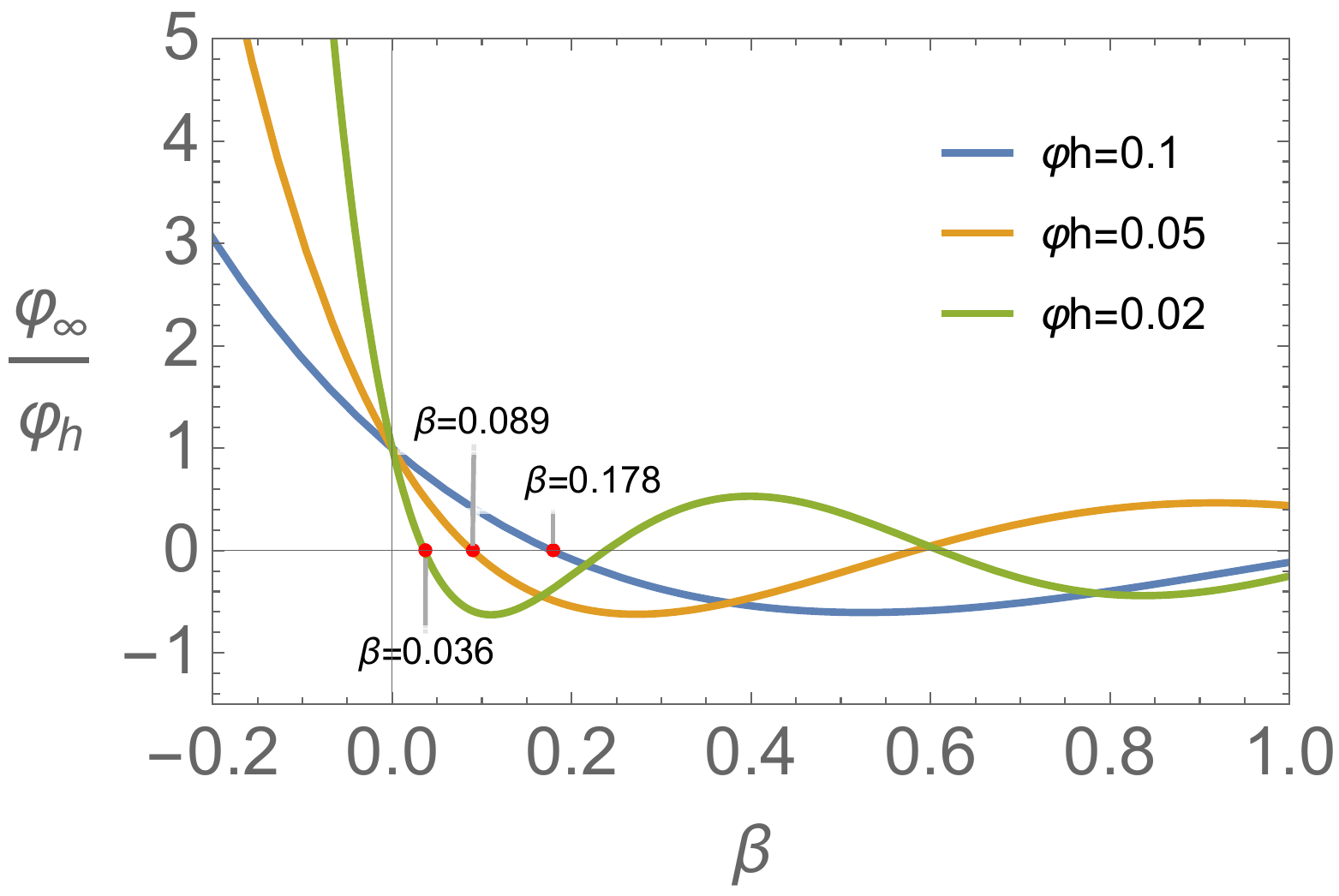}  \includegraphics[scale=0.265]{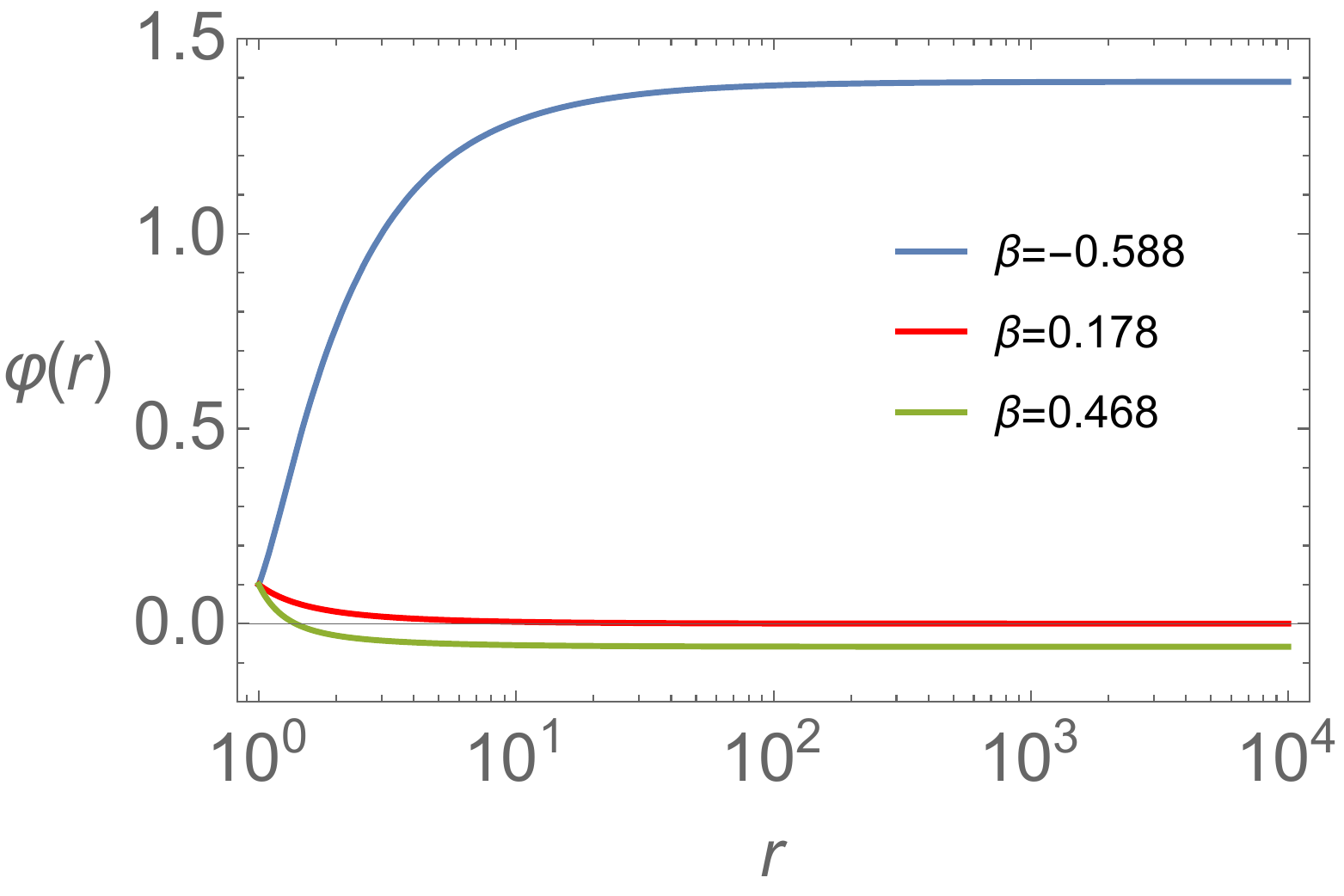} 
\caption{For $f=\alpha \varphi^2$, (left)  $\varphi_{\infty}/\varphi_h$ vs $\beta$ and (right) $\varphi(r)$ for different values of $\beta$ fixing $\varphi_h = 0.1$}\label{fig:ExBInf}

\includegraphics[scale=0.265]{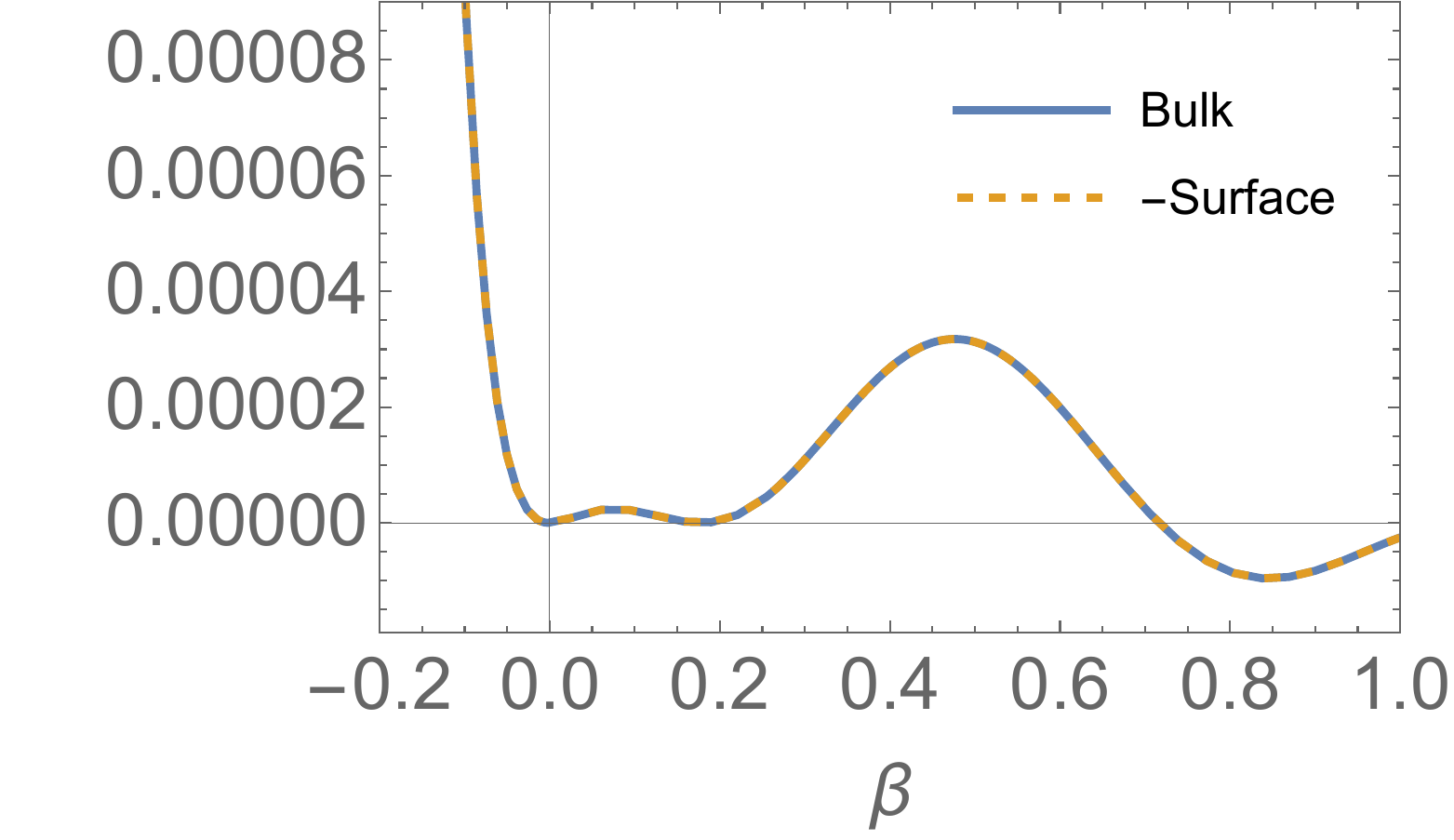} \includegraphics[scale=0.265]{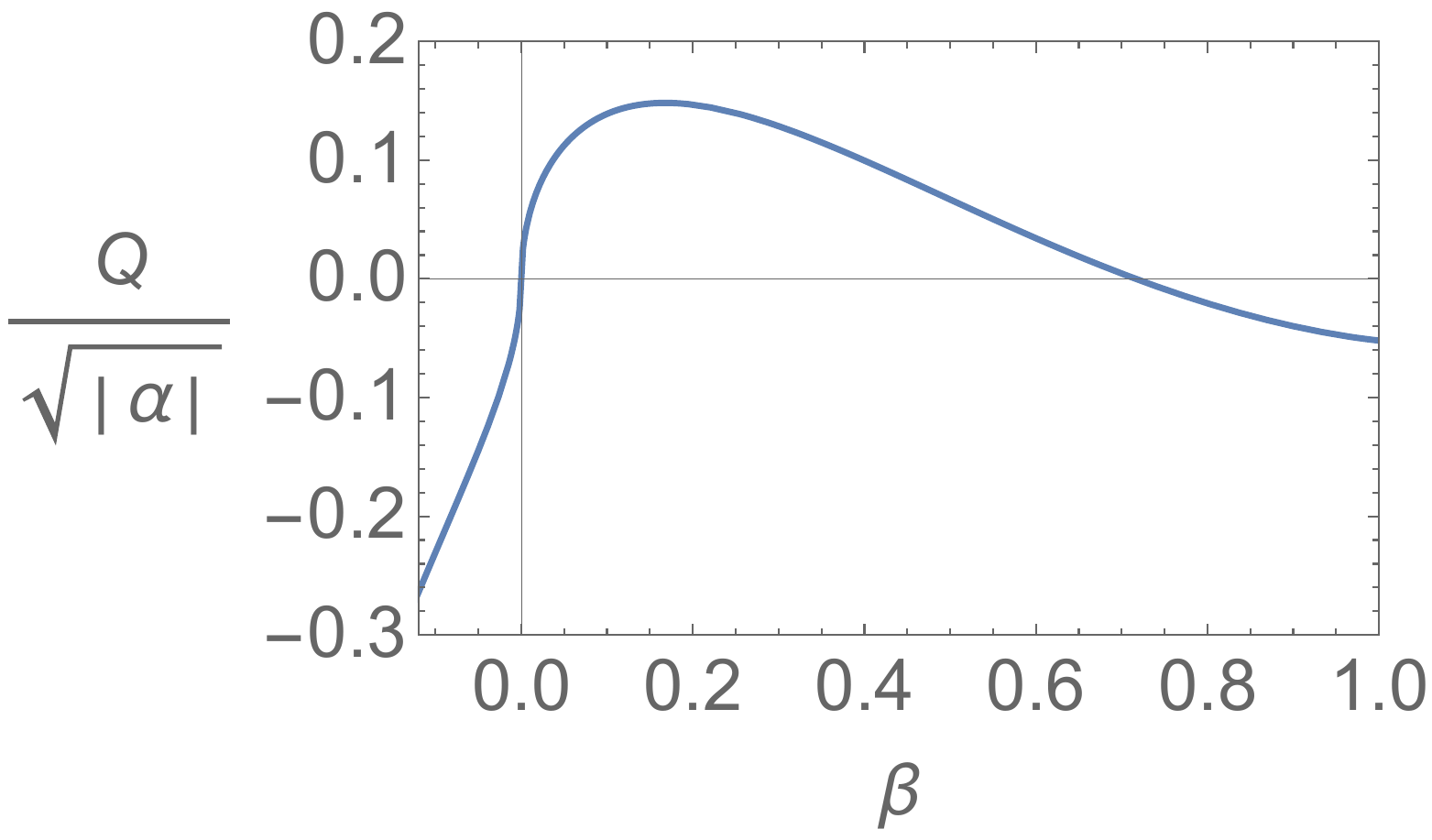} 
\caption{Old no-hair theorem: For $f=\alpha \varphi^2$ with $\varphi_h = 0.1$ (left) plot of bulk and surface term (right) the scalar charge $Q/{\sqrt{|\alpha|}}$ vs $\beta$ } \label{fig:ExBoldNH}
\end{figure}

\begin{figure}[b!]
\includegraphics[scale=0.265]{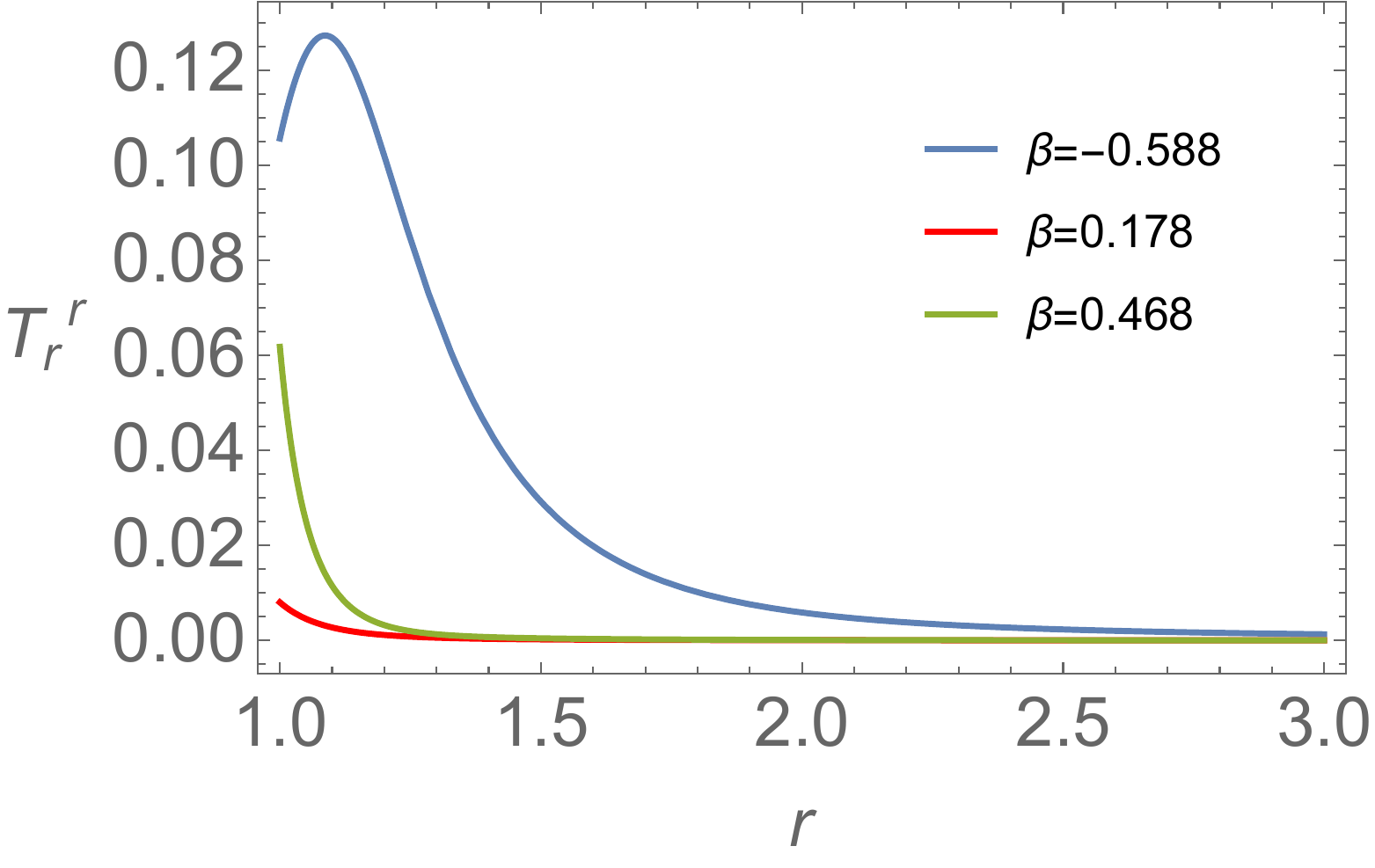}  \includegraphics[scale=0.265]{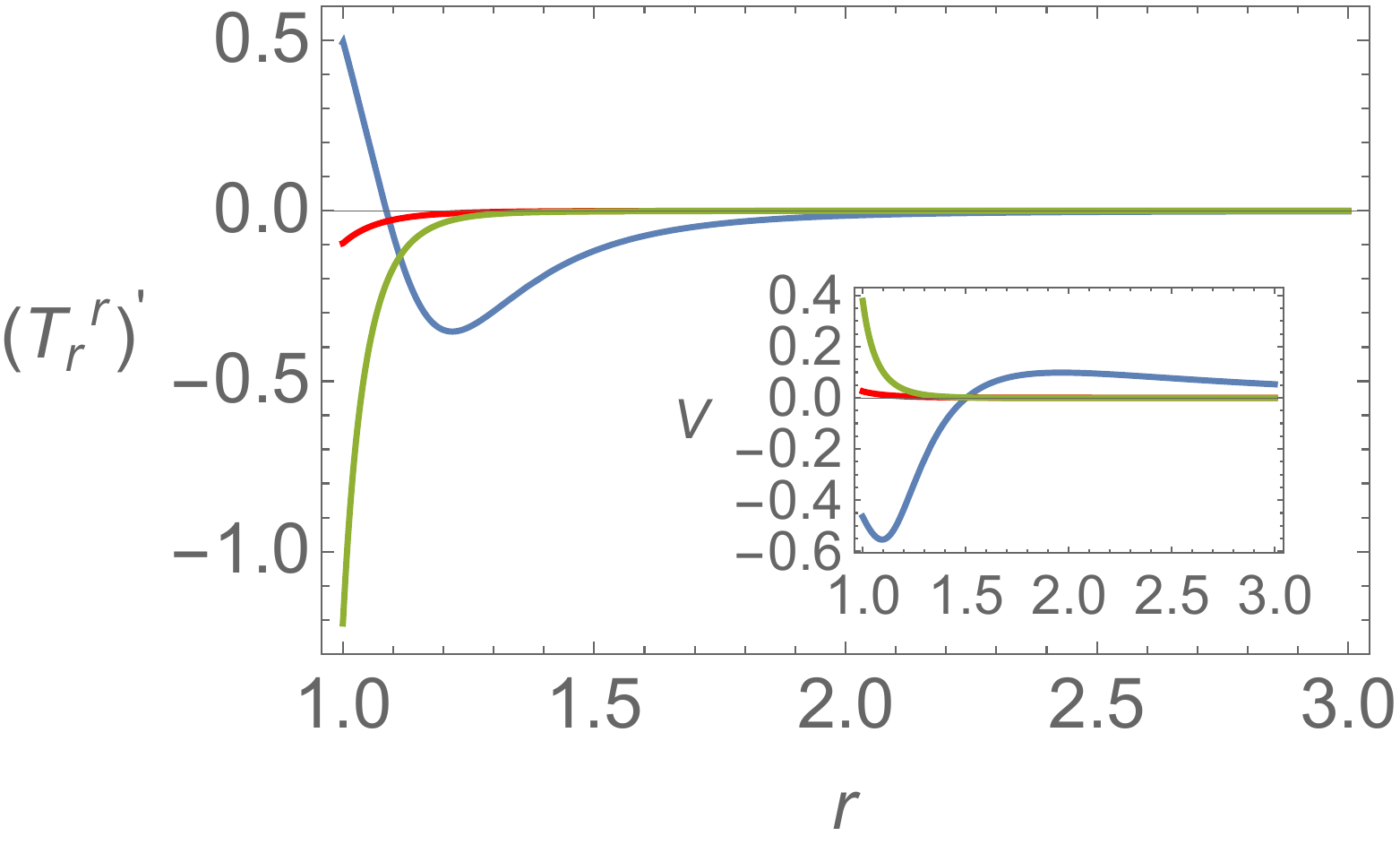}
\caption{Novel no-hair theorem: For $f=\alpha \varphi^2$ and $\varphi_h = 0.1$, (left) $T_r{}^r$ and (right) $(T_r{}^r)'$ and $V$.} \label{fig:ExBTrr}
\end{figure}

In this case, the surface term can vanish when $\varphi_{\infty}=0$ or $\varphi_1 = 0$ which reduces the bulk integration to be zero. This can not be achieved for $f(\varphi)<0$ or $\alpha < 0$ and so the no-hair theorem is expected to hold. This is numerically shown in Fig.\ref{fig:ExBInf}, where the solutions show an ever increasing value of $\varphi_{\infty}$ for negative $\beta$ and therefore certainly there are no solutions that yield $\varphi_{\infty} =0$ for negative values of $\beta$. Moreover, as shown in Fig.\ref{fig:ExBoldNH}, the bulk integration values are growing in the negative $\beta$ region. In addition, when $\varphi_{\infty} \neq 0$ the surface term survives unless $\varphi_1=0$ and the equality (\ref{eq:oldNoH}) becomes non-trivial which indicates the existence of black hole solutions. Thus the old no-hair theorem is expected to be evaded and this is verified by generating numerical solutions in Fig.\ref{fig:ExBInf}. The comparison between the bulk and surface terms are then plotted in Fig.\ref{fig:ExBoldNH} and show that the bulk term is exactly the same as the surface term with the opposite sign. To explicitly demonstrate our argument for the novel no-hair theorem, the energy-momentum tensor $T_{r}{}^r$ and its derivative $(T_{r}{}^r)'$ are depicted in Fig.\ref{fig:ExBTrr}, which shows that $(T_{r}{}^r)'$ can take positive values with $V$ being negative, while smoothly connecting the two asymptotic regimes. This demonstrates that one may obtain solutions for any of the regions displayed in Fig.\ref{fig:novel}.

\section{Conclusion}

We investigated the no-hair theorems in ESGB theory previously studied in \cite{Antoniou:2017acq, Antoniou:2017hxj}. In the formulation of the old no-hair theorem, the surface term has so far been neglected. However, due to the non-minimal coupling, the asymptotic behaviours of the scalar fields are drastically altered and hence the surface term can not be ignored in principle. Consequently, the evasion of the old no-hair theorem should be discussed taking the presence of the surface term into account. We provided the right criteria for the old no-hair theorem to hold or to be evaded. This explains the existence of the numerical solutions when the coupling function is negative, which was excluded in the previous studies \cite{Antoniou:2017acq, Antoniou:2017hxj}, and this fact is demonstrated for the cases of $f=\alpha e^{\gamma \varphi}$ and $f=\alpha \varphi^2$. In the case of the novel no-hair theorem, one generally expects it to be evaded for non-minimal couplings since the original version explicitly assumed a minimal coupling to gravity. We confirm this fact by finding regular solutions numerically, which merely obey the regularity condition (\ref{eq:dvarphi}) on the horizon and we find that it is not necessary for the derivative of the energy-momentum tensor to be negative. Instead, $(T_r{}^{r})'$ may admit either sign on the horizon while still yielding an acceptable solution. In summary, the novel no-hair theorem is evaded automatically in ESGB theory while the old no-hair theorem holds for a very specific choice of parameters and is, in fact, the only way to limit the possible black hole solutions with non-trivial scalar hair.

\vspace{0.5cm}
\acknowledgments
We wish to thank Sang Hui Im for useful discussion. We were supported by the Institute for Basic Science (Grant No. IBS-R018-D1, Grant No. IBS-R018-Y1). We appreciate APCTP for its hospitality during completion of this work.

\vspace{1.2cm}

\bibliography{arxiv}{}
\bibliographystyle{ieeetr}
%
\end{document}